# Holistic Approach to Training of ICT Skilled Educational Personnel

Mariya Shyshkina

Institute of Information Technologies and Learning Tools
of the National Academy of Pedagogical Sciences of Ukraine

`marple@ukr.net`

**Abstract.** The article intends to explore and estimate the possible pedagogical advantages and potential of cloud computing technology with aim to increase organizational level, availability and quality of ICT-based learning tools and resources. Holistic model of a specialist is proposed and the problems of development of a system of methodological and technological support for elaboration of cloud-based learning environment of educational institution are considered.

**Key terms.** KnowledgeEvolution, KnowledgeManagementMethodology, Didactics, KnowledgeManagementProcess, ICTInfrastructure

## 1 Introduction

As it is now impossible to introduce advanced ICT while managing this process without mastering the ICT and other related pedagogical technologies, the main aim is to train ICT-skilled educational personnel. Cloud computing technology (CC) is to create a high-tech learning environment of educational institution, enhancing multiple access and joint use of educational resources at different levels and domains. On this basis it is possible to combine corporate resources of the university and other on-line resources, adapted to learning needs, within a unite framework.

Cloud computing is used for resources supply and to support collaboration in the learning process in particular by means of mobile services. It requires the development of new approaches and models for designing of a learning environment. Among them there are those based on a holistic approach to learning [1], [5], [7], [9].

For this aim a set of instrumentation tools for cloud-based learning resources collection, elaboration and design, holistic models of learning environment and specialist models, and a system of methodological and technological support for the development of cloud-based learning environment of educational institution should be created.

The *purpose of the article* is to identify trends and conceptual models of educational personnel training within the cloud based learning environment.

## 2  Problem Statement

The problem of training of qualified educational management personnel as well as teachers oriented on ICT based learning can nowadays hardly be taken independently from the processes of the innovative development of educational space formed within the school, region and educational system of a country or globally [1]. In this regard, there is a need for fundamental research focusing on the possible ways of developing an educational environment of educational institutions. It should take into account, the trends of improving ICT facilities while searching for new engineering technological decisions and new pedagogical and organizational models [1], [2]. The main focus is on shifting from mass introduction of separate software products, to an integrated and combined environment which supports distributed network services and cross-platform solutions.

Emerging technologies of information and communication networks give a way for implementing a holistic approach to education and training of personnel. A holistic approach focuses on combining science and practice, training and production, fundamental and applied knowledge and technological competencies with social and humanitarian [5], [9]. Above all it aims the development of public administration's management skills in the educational field basing on a unite approach to learning design and management. This is a promising direction for the development of a field's human potential. The innovative processes therefore, of the organization and development of learning environment, search for new approaches and models for specialist education and training becomes a matter of interest [11].

There is **a problem** of availability and valuable ways of learning resources delivery, to achieve with their use the best pedagogical effect and to gain maximum learning potential of ICT. This issue, may be hence partially solved if delivered by means of cloud computing technology [2], [8], [12]. The main advantage of this technology is the improved access to qualitative resources (and sometimes the only possible access to necessary recourses at all). **The idea** is simply to explore approaches for the modeling and estimation of CC-based learning process settings and valuable tools for its organization.

## 3  Education and Training of ICT-Skilled Management and Public Administration Personnel

Public administrators are public servants working in public institutions, departments and agencies [10]. Specifically, they are concerned with "planning, organizing, directing, coordinating, and controlling government operations" [6]. Specific sphere are public servants for education management. For such personnel to be efficiently trained, the need to develop novel approaches arises, as this sphere is mostly concerned with multi-disciplinary knowledge and requires skills on the merge of training, learning and management. Due to the fact that most pedagogical innovations are also based on ICT the need in the sphere of education management also arises. There is a branch of pedagogical sciences dealing with theoretical and methodological problems of ICT in education use, psychological and pedagogical substantiation of these processes, elaboration of ICT tools and resources for providing functioning and development of educational systems. So there should be specialized personnel to insure the processes of implementation, introduction and development of ICT-based learning technologies within the sector of public administration.

There are significant needs in IT competent specialists in the sphere of public administration. Without ICT competence or competence in ICT for learning, problems with their adaptation at the workplace arise, as do problems with the necessity of additional and often profound training almost immediately after hiring. In some cases, a vague idea of future graduates about the real problems and conditions of work with innovative ICT infrastructures and ICT-based tools leads to lack of commitment to practical solutions of work situations thus to a low level of innovative inclusion.

Formation of the innovative institution's ICT infrastructure could solve some of the aforementioned problems [11]. Namely, it would bridge the gap between the process of training and the level of demand for their product. An environment that would bring together the learning resources of educational and industrial projects would be created, and would cover different levels of training; including the training of both students and pedagogical management personnel.

According therefore to the high rates of development of both the global ICT market for the education sector, and the IT market of learning tools, the problem of training professional staff for the domestic public administration and the IT-oriented sector of education management; personnel which are primarily prepared within higher and post graduate schools (e.g. universities and advanced training schools) being continuous, we conclude that modern approaches to the design of educational systems are a key point.

It is unlikely that the current state of skilled personnel of management and public administration of education could be regarded as fully satisfactory for the needs of innovative development of ICT-based learning, for the required number of qualified professionals with appropriate structure and quality of training. The system of training and retraining of employees for public administration has not been properly formed.

These problems should be considered within the context of development of an institution's and a region's innovative environment as well as on national and international level [3], [11]. These processes have to do with the modernization of a learning environment in perspective of emerging ICT. Thus the developmental need of new models and approaches to personnel training arises, which will account for the modernization of ICT infrastructure and integrate resources of different levels and use.

Introduction of innovations into the educational environment of a state or a region, is highly concerned with the development of human resources of informatization on education [1]. It requires new types of skills and competencies which graduates often lack of. These skills include leadership, ability to approach a problem holistically, and the ability to critically evaluate achievement and self-assessment [9], [11]. It is the lack of qualified personnel and the absence of a strategic approach to ICT infrastruc-ture design that are among the reasons for an institution's of professional education deficiency of a unite high-tech desisions.

Nowadays content-technological process regarding the creation and use of ICT products, and in particular the electronic learning resources, requires fundamental background knowledge in both ICT and pedagogy. The approaches however for training personnel today, do not sufficiently take into account the recent years' innovative changes in the ICT industry, nor the real needs regarding the extent of such training.

A mean for provision of users with relevant services of cloud computing technology is considered to be outsourcing; i.e. a service in a specific system to implement its core functions is required, offered and sold by another system external to this [3]. ICT outsourcing plays an important role in enhancing the scientific and technical level of ICT-systems of an educational institution as well as the efficiency of their operation and their development. It is a market mechanism incorporating the latest advances in the ICT sector and to satisfying user demand [2], [3].

The main problem in educational practice is the contradiction between on one hand the objective need for a continuous improvement of the software and the hardware power of training computer complexes, and on the other, the lack of personnel's ability (in both qualitative and quantitative manners) to maintain, manage and develop their ICT systems appropriately. The informatization hence of an educational institution in terms of cloud computing and ICT outsourcing, will offer realistic solutions for both the deepening of informatization and improvement of ICT's educational performance and use of information resources [2], [3].

The basic principles of such introduction should be: a tight relationship of learning with training and methodological support for tutors, focus on a specific educational task; modularity of learning; continuity of learning, sharing experience and formation and participation in professional association activities (including electronic) [2], [3]. In this process electronic distance learning systems should be actively used, based on the principles of open education, with the maximum possible use being of CC technology and outsourcing.

## 4    What are Advantages of Cloud Computing Decision?

### 4.1    It is a Cost Effective Solution

Being cost-effective the user can get (buy) products and services proposed by the virtual supermarket of ICT according to their needs (individual or group, collective, corporate), they may pay only for what has been bought (e -transport, e-content, e-services, virtual e-tools, a generic and subject software applications, network platforms - full range of cloud services along with services for the design and implementation of ICT systems and their fragments ordered by the users, their warranty and post warranty service, maintain, upgrade and improvement, etc.) and only for the actual time of use of the purchased product [3 ]. This will allow users to avoid regular updating and upgrading of powerful general system software and hardware tools of their own ICT systems, avoiding a potential surplus of ICT products used from time to time; fragmentary, not fully, as well as spare parts, reduction of requirements for information security of their own ICT systems, reduction of the number of their ICT services and requirements for professional competence of their employees and as a result, significantly reduce overall costs to support the operation and develop their ICT systems, to increase their social and economic return, their efficiency [2], [3].

### 4.2   This is a Flexible Solution of ICT Infrastructure

It is designed for increased flexibility and effective access to learning resources so as to build a unified and mobile infrastructure.

On the basis of CC infrastructure all main aspects of interaction of a learner may be comprehended on the unite basis. Along to approach introduced in [1], among them there will be interactions between a learner and other learners; a learner and a teacher; a learner and a learning tool; a learner and educational institution; a learner and the society. This will lead to an environment of learning organization on the unite base, where collaboration between learners and a tutor, free and flexible resource access, learning activity within social inclusion into the environment of an educational institution and the society will be enabled. The ICT support of learning is realized by means of cloud services. It is designed for adaptation to the rapidly changing external/internal environment, changing of task/competence requirements and development of modern pedagogical approaches.

Due to the principles of open education [1], there is a need to create an innovative learning environment that will form and develop necessary professional skills. Among them are leader skills, collaborative skills, critical thinking, and the ability to view a problem in a holistic manner. These skills refer mostly to the demand of the sphere of public administration of education, as in alliance with them; a process of innovative development may be involved. This may be achieved on the basis of a holistic approach to specialist training when the planning, design and resource management and learning activity of an organization and its monitoring, may be represented on a unite basis. It will be achieved through the unite development of different competencies: professional, fundamental, personal and technological.

## 5   A Holistic Model of a Specialist

Holistic approach to education deals with the learning processes to be taken as unity of all main aspects of a personality development, for example such as mental, emotional and volitional. This is in tune with a meaning of the term "holistic" as completeness, being impossible with disregard of some of its components.

There are innumerous investigations devoted to the problems of holistic learning development in different aspects such as learning and teaching interaction, collaboration processes, engagement of both aspects of theory and practice to gain comprehensive view of a subject [7], [9]. Now there are important trends of research development in concern to modern ICT. For example, holistic view is to approach learning environment structure. Thus, the model of a learning environment, developed in [1] is to reveal main components and types of interactions within the different learning process settings.

The notion of holistic learning occurs in relation to personnel training, concerning to different components and interactions within educational organization. It may touch upon certain types of activity, collaboration and resource management processes, engaging thus the entire organization at all levels and developing a performance culture of personnel. There are different ways to approach peculiarities of specialist formation, namely in the aspect of personal or professional features. That concerning to modeling of professional competencies [5], especially in the sphere of educational management. Another aspect is about holistic models to develop leader skills [4], [9], which are more to traits of a personality.

The proposed approach is based on holistic model of a specialist in the sphere of informatization of education presented in Fig.1. It concerns to Domain Competencies which would occupy fundamental knowledge of educational management and modern learning technologies and also ICT skills and ability to use e-learning tools. There are also Personal Competencies, such as leader skills, critical thinking, and capability to holistic view of a problem, responsibility and activity of an individual. As for professional skills there are planning, design, resources management, cooperation and collaboration skills, performance skills and ability for monitoring and self evaluation.

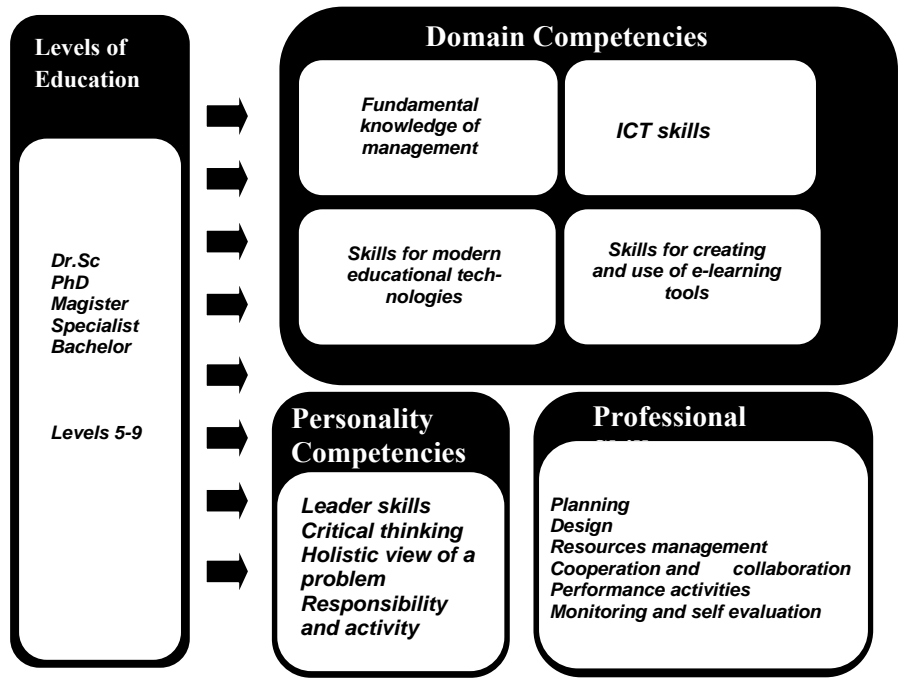

**Fig. 1.** A holistic model of a specialist

All the components of a specialist's competencies, skills and knowledge are consistently formed within the main level of education which corresponds to National qualification framework (levels 5-9).

Cloud computing decision is a reasonable way to support holistic learning settings giving a platform for unite representation and access to learning tools for different levels and domain of education as also for different individuals and groups of users.

*Promising ways of assessing resources quality, while building a holistic learning environment are:*

A. Analysis of the most appropriate ways to use cloud computing technology to supplementing and structuring collection of educational learning resources, filling it with the resources on this basis and organizing multiple access to their use
B. Use of a certain set of educational resources for testing methods to evaluate the quality of their use within the cloud-based infrastructure of organization
C. Recommendations on methods to replenish the collection, its prototyping and ways of structuring resources
D. Elaboration of requirements to provide electronic resources, for collection replenishment
E. Analysis of cloud computing technology outsourcing for optimal selection and use of resources' collections
F. Creation of recommendations to developers and material for replenishment and application of existing electronic learning resources
G. Development of recommendations for dissemination and use of collections of electronic resources

## 6   An Expected Impact and Social Results of the Project

The important step to wider application and introduction of new learning approaches and to gain most possible benefit from emerging technologies and ICT tools should be achieved through modernization and upgrading of ICT learning environment of educational institutions, increasing of overall level of e-learning.

To achieve these goals the main problem is to rise ICT and professional level of competencies of subjects of the learning process – managers, pedagogical personnel and staff and also personal of ICT departments. Just the people are the most valuable factor of empowerment of development and formation of social and economical systems and educational systems in particular. Just the people are the most important resource which should be involved so as to improve the quality of these social systems and to manage their purposeful and productive growth. By this reason development of tools and resources to train teachers and stuff is critical point because it really concern to all levels of educational systems functioning.

The whole impact of implementation of learning tools and techniques based on cloud computing is aimed at:
- Broaden use of ICT in education aiming at wider take up by learners and teachers
- Effective public-private partnerships for introduction and managements of learning environment solutions

- More efficient introduction of ICT into the learning process through the exploitation of monitoring and assessment tools
- More timely and purposeful acquisition of skills and competences through ICT-based learning technologies, in educational establishments and public administrations
- Increased involvement with the adoption of learning digital technologies

The important step to wider application and introduction of new learning approaches and to gain most possible benefit from emerging technologies and ICT tools should be achieved through modernization and upgrading of ICT learning environment of educational institutions, development of new learning approaches, creating more advanced learning technologies [1], [2].

Formation of innovative ICT infrastructure of the institution could solve some of the problems of development highly skilled educational and management personnel, bridging the gap between the process of training and the level of demand for their product.

Due to development of cloud computing technologies opportunities, functionality and access to collections of electronic learning resources has significantly increased. In this regard, cloud computing is a promising direction of development of electronic resources' collections (may be relevant for development of collections), as it allows the creation of a unified methodology for a single platform, a framework for development and testing, and for improvement and elaboration of integrated assessment methods' quality. This gives an added value to available recourses [2], [11].

The **social results** will help to modernize the learning environment of educational institutions and organizations, to increase educational potential of ICT and add value to the best examples of available learning resources due to their flexible and learner-adaptive access.

At the same time there are several aspects of the cloud-based learning architecture to be a subjected to further research. There are problems of pedagogical and psychological support in regard to the processes of the design and organization of an educational institution's cloud infrastructure, prospecting possible organizational structures to provide learning environment functioning and to teach educational managers and organizers, pedagogical and technical stuff how to use new methods and approaches to learning, based on cloud computing.  There is a necessity therefore, to create an educational and training system of support used by management personnel, teachers and learners.

The result of instrumentation for cloud-based learning resources collection elaboration, and development of cloud-based learning environment of educational institution might be used within different learning and organizational educational structures.

## 7   Analysis and Estimation of Perspective Ways of Development

The cloud based learning infrastructure is to give the opportunities:
- To combine the processes of development and use of electronic resources to support learner competencies

- To insure holistic approach to specialist education and training, combining both technological and social competences, development of critical skills of a learner
- To integrate the processes of training, retraining and advanced training, at different levels of education by providing access to electronic resources of a unite learning environment
- To solve or significantly mitigate the problems of association of electronic resources of the institution into unite framework
- To access to the best examples of electronic resources and services to those units or institutions, where there is no strong ICT support services for e-learning
- To provide of invariant access to learning resources within the unified educational environment, depending on the purpose of study or educational level of the student, enabling person-oriented approach to learning
- To make conditions for a higher level of harmonization, standardization and quality of electronic resources, which may lead to emergence of the better examples of learning resources and to more massive use them

## 8    Conclusion

There are real advantages of CC technologies to assure more flexible, scalable and cost-effective decisions of access to learning resources as within the learning environment of the university and also in learning environment of the whole region, national and international scale. This is an advantage so as to ensure joint use and widening participation in the learning courses of learners from different institution were necessary services are substantiated and supported. As if holistic approaches to cloud services development are already used in education so the challenge is to transfer this experience into wider context.

The project is implemented within the framework of the joint research laboratory of Cloud computing in education of the Institute of Information Technologies and Learning Tools of NAPS of Ukraine (Kiev) and the Krivoy Rog State University (Krivoy Rog), www.ccelab.ho.ua.